\documentclass[twocolumn,floats,showpacs,prl]{revtex4}
\usepackage[above]{placeins}
\usepackage{amsmath}
\usepackage{graphicx}

\begin{document}

\title{Theory of superconducting and nonsuperconducting stripe phases in the
hole-doped superconductor Ca$_{2-x}$Na$_{x}$CuO$_{2}$Cl$_{2}$ ($x=1/8,1/16$)}
\author{Xiuqing Huang$^{1,2}$}
\email{xqhuang@nju.edu.cn}
\affiliation{$^1$Department of Physics and National Laboratory of Solid State
Microstructure, Nanjing University, Nanjing 210093, China \\
$^{2}$ Department of Telecommunications Engineering ICE, PLAUST, Nanjing
210016, China}
\date{\today}

\begin{abstract}
{Based on the newly developed real-space spin-parallel pairing and
superconducting theory, we explore a simple explanation for the
observed checkerboard patterns in hole-doped
Ca$_{2-x}$Na$_{x}$CuO$_{2}$Cl$_{2}$ cuprate superconductor. At a
hole concentration $x=1/8=0.125$, the analytical results show that
there exists a phase competition between non-superconducting
$4a\times 4b$ ($a=b$) checkerboard phase and $8a\times 2c $
superconducting vortex phase, where $a$, $b$ and $c$ are the lattice
constants of the superconductor. At a lower hole concentration $%
x=1/16=0.0625 $, it is revealed that the metastable $4\sqrt{2}a\times 4\sqrt{%
2}a$ checkerboard phase can reorganize itself into a more stable octahedron
phase with the $4a\times 4a$ checkerboard symmetry.}
\end{abstract}

\maketitle

The hole-doped Ca$_{2-x}$Na$_{x}$CuO$_{2}$Cl$_{2}$ (CNCOC), with a crystal
structure similar to that of La$_{2-x}$Sr$_{x}$CuO$_{4}$ (LSCO), has been
studied extensively because of the simplicity of its structure. Single
crystals of lightly doped CNCOC can be cleaved easily like mica, which
provides an ideal surface for scanning tunneling microscopy or spectroscopy
(STM and STS) studies \cite{hanaguri,kohsaka}. At the doping level 1/8,
results of STM and angle-resolved photoemission spectroscopy (ARPES)
experiments on CNCOC suggest a checkerboard-like spatial modulation of
electronic density of states with a periodicity of $4a\times 4a$ \cite%
{hanaguri,shen}, which has been shown to be a universal feature of cuprate
superconductors \cite{tranquada,vershinin}. Many theoretical efforts have
been made to explain the non-dispersive checkerboard charge ordering
patterns \cite{hanaguri,komiya,altman,anderson2,hdchen1,tesanovic,kivelson}.
Even though the checkerboard pattern can be simulated numerically by
adjusting different parameters in a number of proposed models, the exact
causes of the checkerboard pattern in cuprate superconductors are still not
conclusive.

Recently, we have proposed a real space spin-parallel mechanism \cite%
{huang1,huang2,huang3} of superconductivity which has successfully provided
coherent explanations to a number of complicated problems in conventional
and non-conventional superconductors (including the new iron-based materials
\cite{kamihara,xfchen}). Our work marks an important step forward in
unraveling in the mystery of the superconductivity. In the present paper, we
will show that our simple pictures of Cooper pairing and vortex lattices can
lead to new understanding of the emergence of non-superconducting
checkerboard phases and superconducting vortex phases in CNCOC
superconductor.

\begin{figure}[tbp]
\begin{center}
\resizebox{1\columnwidth}{!}{
\includegraphics{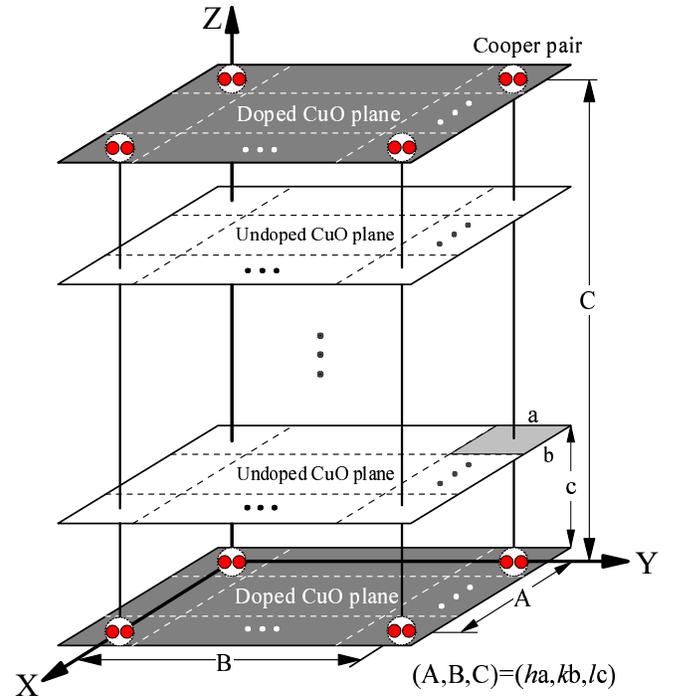}}
\end{center}
\caption{{}\textbf{Simplified schematic unitcell of the
electron-pairs (Cooper pairs) Wigner crystal in the
high-}$T_{c}$\textbf{\ cuprates.}} \label{fig1}
\end{figure}

It is well known that the formation of stripe patterns is generally
attributed to the competition between short-range attractive forces
and long-range repulsive forces \cite{seul}. In a superconductor
with the primitive cell ($a,b,c$), at a rather low doping level, the
interactions among electrons can be neglected and the superconductor
behaves much like a charged random system. As more carriers are
added, the effect of the competitive interactions among electrons
will emerge. At a proper doping level (not too low, not too high),
the electron pairs can self-organize into a low-temperature
orthorhombic (LTO) phase (Wigner crystal of Cooper pairs) of
$(A,B,C)=(ha,kb,lc)$, as shown in Fig. \ref{fig1}. Thus, the carrier
density $x$ is given by

\begin{figure}[tbp]
\begin{center}
\resizebox{1\columnwidth}{!}{
\includegraphics{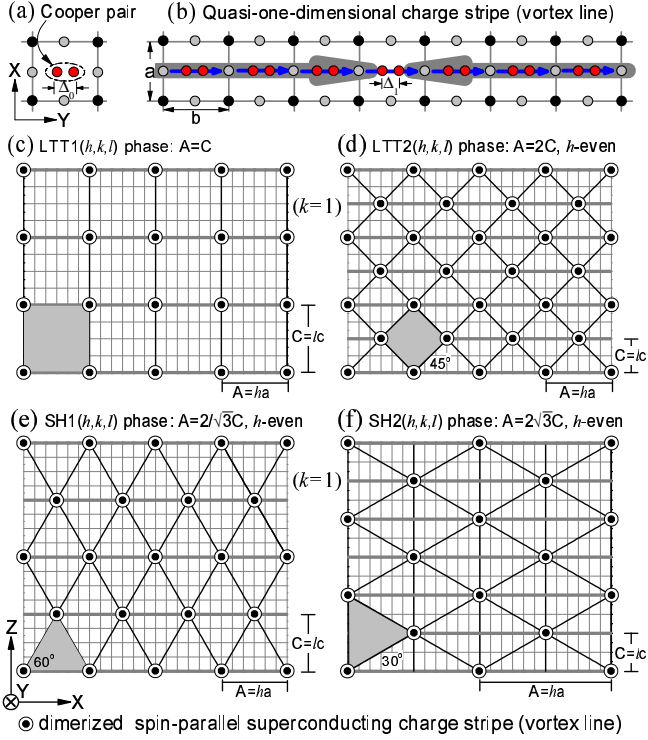}}
\end{center}
\caption{\textbf{The schematic interpretation of the theory of
superconductivity based on the minimum energy principle in cuprate
superconductors. (a),} A quasi-zero-dimensional localized Cooper
pair (with the electron- electron separation $\Delta _{0}$) located
inside a square lattice of two-dimensional CuO plane. \textbf{(b),}
A quasi-one-dimensional dimerized vortex line and a collective
real-space spin-parallel confinement picture in CuO plane, where
$\Delta _{1}$ ($<\Delta _{0}$) is the electron- electron separation.
\textbf{(c-f), }Four quasi-two-dimensional vortex
lattices with a uniform distribution of vortex lines. \textbf{(c),} LTT1($%
h,k,l $) phase, the charge stripes have a tetragonal symmetry in XZ
plane. \textbf{(d),} LTT2($h,k,l$), the vortex lattice has a
tetragonal symmetry in XZ plane with a orientation 45$^{0}$.
\textbf{(e)} and \textbf{(f),} The simple
hexagonal (SH) phases [SH1($h,k,l$) and SH2($h,k,l$)]. Where $\protect\xi %
_{xz}$ is the nearest neighbor stripe-stripe separation.}
\label{fig2}
\end{figure}

\begin{equation}
x=p(h,k,l)=2\frac{V_{abc}}{V_{ABC}}=2\times \frac{1}{h}\times \frac{1}{k}%
\times \frac{1}{l},  \label{fractions}
\end{equation}%
and the corresponding charge carrier density is%
\begin{equation}
\rho _{s}=\frac{2}{ABC}=\frac{2}{hkl}\frac{1}{abc}=\frac{x}{abc},
\label{density}
\end{equation}%
where $h$, $k$, and $l$ are integral numbers, and $V_{abc}$ and $V_{ABC}$\
are the unit cell volumes of the\ lattice and the corresponding
superlattice, respectively.

Physically, electron pairing (Cooper pair) in cuprates is an individual
behavior characterized by pseudogap, while superconductivity is a collective
behavior of many coherent electron pairs. To maintain a stable
superconducting phase (minimum energy), first the Cooper pairs of Fig. \ref%
{fig2}a must condense themselves into a real-space quasi-one-dimensional
dimerized vortex line (a charge-Peierls dimerized transition), or a Cooper
pairs's charge river in the CuO$_{2}$ plane of cuprate superconductor, as
shown in Fig. \ref{fig2}b. It should be noted that this figure also
illustrates a collective real-space spin-parallel confinement picture where
any electron pair inside always experiences a pair of compression forces
(indicated by the two big arrows). And second, in order to further minimize
the system energy, the vortex lines must self-organize into four possible
quasi-two-dimensional vortex lattices where a uniform distribution of vortex
lines is formed in the plane perpendicular to the stripes, as shown in Figs. %
\ref{fig2}c-f. In the picture,\ superconducting charge stripe and the vortex
line are exactly the same thing. Moreover, this scenario indicates that the
superconductivity is relevant to the lattice constants, in support of a
recent experiment which has shown that cuprate superconductivity can be
varied by the interatomic distances within individual crystal unit cells
\cite{slezak}.

\begin{figure}[tbp]
\begin{center}
\resizebox{1\columnwidth}{!}{
\includegraphics{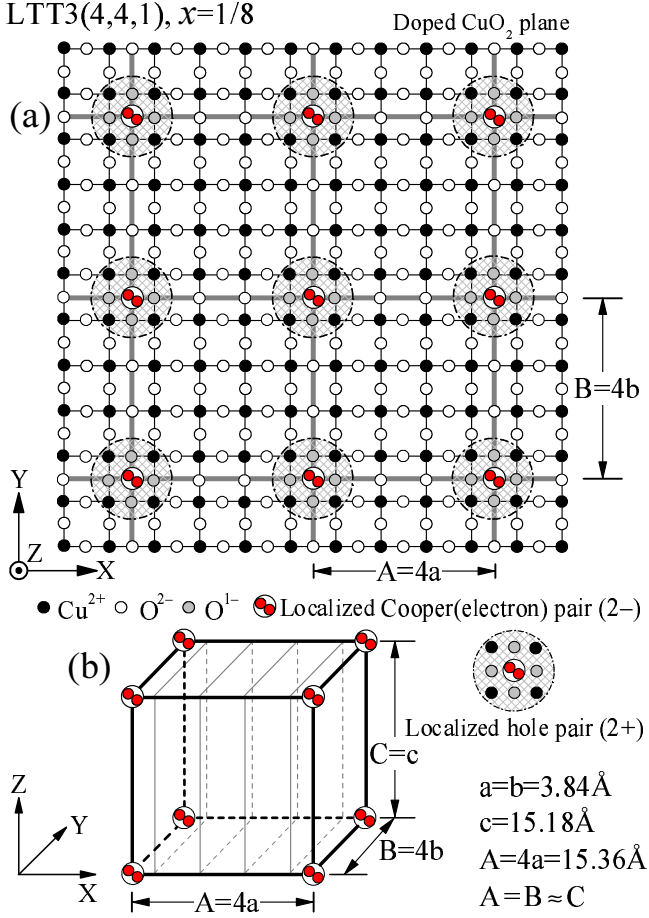}}
\end{center}
\caption{\textbf{The nondispersive LTT3(}$4,4,1$\textbf{)
superlattice of the localized electron pairs in CNCOC at doping
level }$x=1/8$\textbf{.} \textbf{(a),} The $4a\times 4a$
checkerboard in the doped CuO$_{2}$ plane. \textbf{(b),} The
localized Cooper pairs form a non-superconducting stabilized
simple-cubic structure ($A=B\approx C$).} \label{fig3}
\end{figure}

In the LTT1($h,k,l$) phase, as shown in Fig. \ref{fig2}c, the charge stripes
have a tetragonal symmetry in XZ plane in which the superlattice constants
satisfy
\begin{equation}
\frac{A}{C}=\frac{ha}{lc}=1.
\end{equation}%
Fig. \ref{fig2}d shows the LTT2($h,k,l$), the vortex lattice has a
tetragonal symmetry in XZ plane with a orientation 45$^{\text{0}}$ and the
superlattice constants:
\begin{equation}
\frac{A}{C}=\frac{ha}{lc}=2.
\end{equation}

While in simple hexagonal (SH) phases, as shown in Figs. \ref{fig2}e and f,
the charge stripes possess identical trigonal crystal structures. In the SH1(%
$h,k,l$) phase [see Fig. \ref{fig2}e], the superlattice constants have the
following relation
\begin{equation}
\frac{A}{C}=\frac{ha}{lc}=\frac{2\sqrt{3}}{3}\approx 1.15470.
\end{equation}%
For the SH2($h,k,l$) phase of Fig. \ref{fig2}f, this relation is given by
\begin{equation}
\frac{A}{C}=\frac{ha}{lc}=2\sqrt{3}\approx 3.46410.  \label{lattice}
\end{equation}

\begin{figure}[tbp]
\begin{center}
\resizebox{1\columnwidth}{!}{
\includegraphics{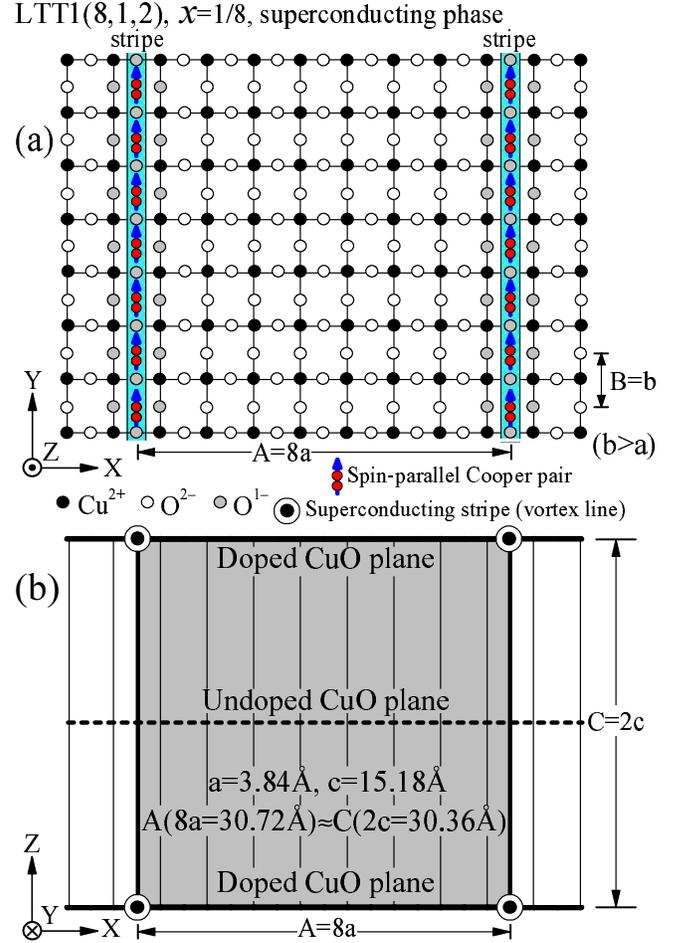}}
\end{center}
\caption{\textbf{The LTT1}$(8,2,1)$\textbf{\ superconducting phase of }$%
x=1/8 $\textbf{\ sample for CNCOC with }$A=8a\approx C=2c.$\textbf{\
(a),} The doped CuO$_{2}$ plane with a uniform stripe spacing
$A=8a$. \textbf{(b),} The tetragon vortex structure ($A=C$) where
only half of the CuO$_{2}$ planes are doped. } \label{fig4}
\end{figure}

It is worth to emphasize that our theory of Fig. \ref{fig2} is based on the
most solid \textit{minimum energy principle}. Physically, in a material, the
dominant structural phase should be a minimum-energy state which satisfies
the basic symmetry of the crystal structure. In this sense, the
superconducting states are merely some minimum energy condensed states of
the electronic charge carriers, or some kinds of real-space low-energy
Wigner-crystal-type charge orders. We argued that the appearance of the
stable vortex lattices (see Fig. \ref{fig2})\ is a common feature of the
optimally doped superconducting phases \cite{huang3}. But, for non-optimal
doping samples we found that the vortex lattices tend to form the
superconducting low-temperature orthorhombic (LTO) phase where the
superlattice constants satisfy $A\neq B\neq C.$

We consider the lattice-constant-dependent schematic [Figs. \ref{fig1}$-$\ref%
{fig2} and Eqs. (\ref{fractions})$-$(\ref{lattice})] a promising approach to
the checkerboard problem in CNCOC, as it can naturally explain both the
longstanding puzzle of \textquotedblleft magic doping
fractions\textquotedblright\ and checkerboard pattern in LSCO \cite{huang1}.
\ In CNCOC, the experimental lattice constants are $a\approx b=3.84\mathring{%
A}$ and $c=15.18\mathring{A}$. From Eq. (\ref{fractions}), it is clear that
the nondispersive superlattices of $4a\times 4a$ in CuO$_{2}$ planes can be
expected at $x=1/8$ of phase LTT3($4,4,1$) with $A=B$ (or $p(4,4,1)$ phase),
as shown in Fig. \ref{fig3}a. From the structure parameters, one has $%
A=B=4a(\sim 15.36\mathring{A})\approx C=c(\sim 15.18\mathring{A})$, this
implies that it is possible for the localized Cooper pairs in CNCOC to form
a non-superconducting stabilized simple-cubic structure at $x=1/8$, as shown
in Fig. \ref{fig3}b. Furthermore, the real space pictures of
\textquotedblleft localized hole pair\textquotedblright\ and
\textquotedblleft localized electron pair\textquotedblright\ are illustrated
in the figure.

\begin{figure}[tbp]
\begin{center}
\resizebox{1\columnwidth}{!}{
\includegraphics{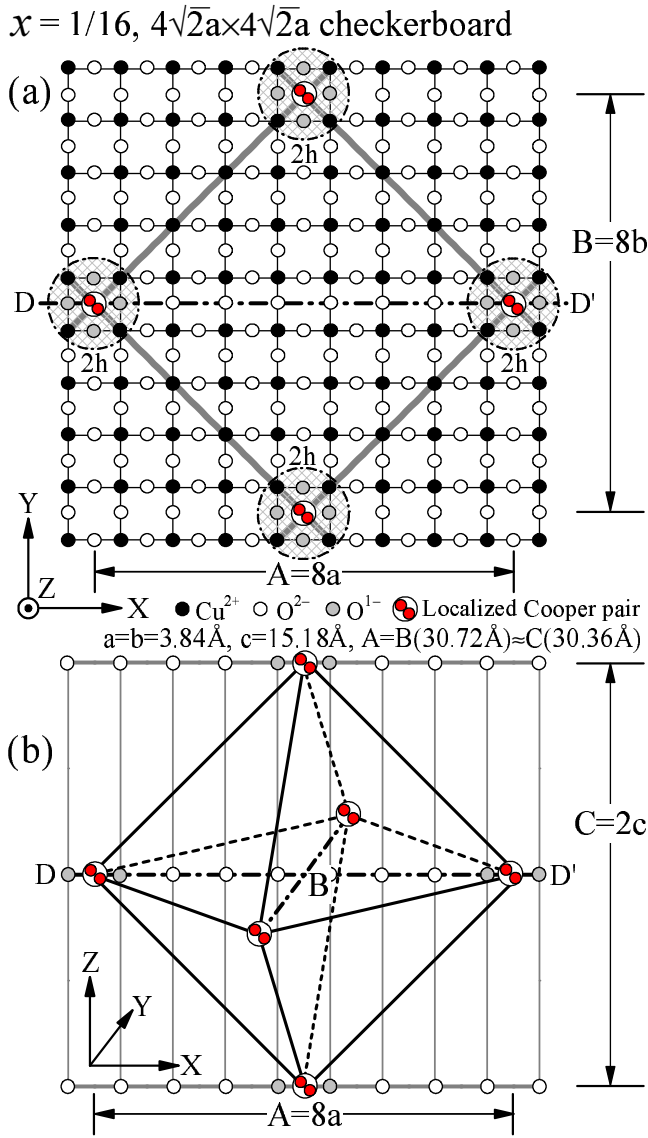}}
\end{center}
\caption{\textbf{The nondispersive superlattices of the electron
pairs in
the CNCOC at doping level }$x=1/16$\textbf{. (a),} The two-dimensional $4%
\protect\sqrt{2}a\times 4\protect\sqrt{2}a$ checkerboard in the doped CuO$%
_{2}$ plane. \textbf{(b),} The three-dimensional stable octahedron
superlattice of Cooper pairs.} \label{fig5}
\end{figure}

Since $A(\sim 30.72\mathring{A})\approx C(\sim 30.36\mathring{A})$, the $%
x=1/8$ sample of CNCOC may also be possible in the LTT1$(8,2,1)$
superconducting phase of Fig. \ref{fig2}c, where the vortex lines are formed
in CuO XY-planes with a uniform spacing of $A=8a$ while the low-temperature
tetragonal vortex lattice is established in XZ plane, as shown in Fig. \ref%
{fig4}a and Fig. \ref{fig4}b, respectively. The formation of this
superconducting phase is in competition with the predominant
non-superconducting phase of Fig. \ref{fig3}.

Most recently, a theoretical prediction of the checkerboard pattern has been
carried out with the solution of the $4\sqrt{2}a\times 4\sqrt{2}a$
superstructure in CNCOC superconductor at $x=1/16$ \cite{patterson}.
However, it has been found that two different LSCO compounds ($x=1/8$ and $%
x=1/16$) can exhibit the same nondispersive $4a\times 4a$ superstructure
within their CuO$_{2}$ planes \cite{zhou2}. Therefore, these results raise
an important question: whether or not the $4\sqrt{2}a\times 4\sqrt{2}a$
superstructure indeed represents a physical reality in CNCOC.

\begin{figure}[tp]
\begin{center}
\resizebox{1\columnwidth}{!}{
\includegraphics{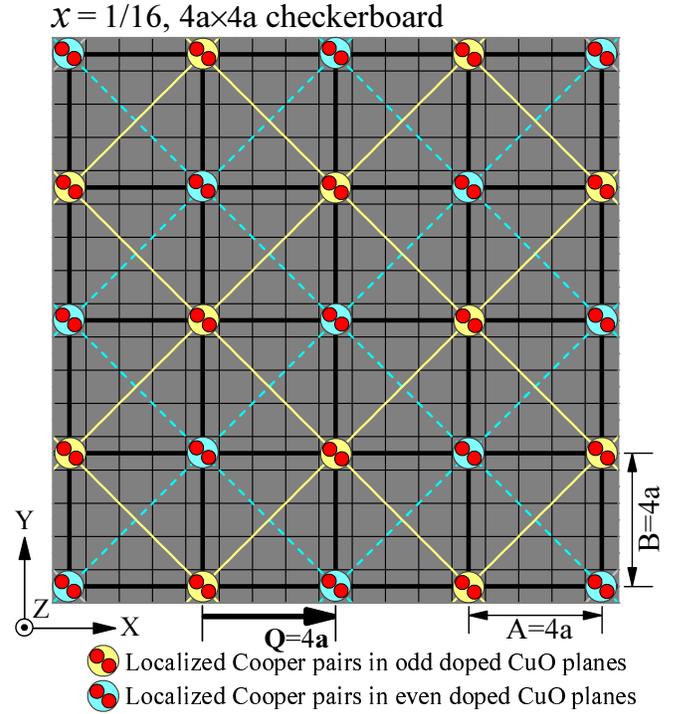}}
\end{center}
\caption{\textbf{A global }$4a\times 4a$\textbf{\ checkerboard
pattern at
x=1/16 sample of CNCOC.} The doped CuO$_{2}$ planes of Fig. \protect\ref%
{fig5} can be divided into \textquotedblleft odd doped CuO
planes\textquotedblright\ and \textquotedblleft even doped CuO
planes\textquotedblright\ and there is a displacement
$\mathbf{Q}=4\mathbf{a} $ between them. This figure shows that
though a single CuO$_{2}$ exhibits
the $4\protect\sqrt{2}a\times 4\protect\sqrt{2}a$ checkerboard pattern at $%
x=1/16$, the results of STM experiment on this sample may still be the $%
4a\times 4a$ pattern.} \label{fig6}
\end{figure}

According to our theory, the localized Cooper pairs in the doped CuO$_{2}$
of CNCOC can exhibit the $4\sqrt{2}a\times 4\sqrt{2}a$ checkerboard pattern
at $x=1/16$, as shown in Fig. \ref{fig5}a. Due to the fact that $A=B(30.72%
\mathring{A})\approx C(30.36\mathring{A})$, the localized Cooper pairs have
a strong tendency to form the most stable octahedron superlattice structure,
as shown in Fig. \ref{fig5}b. In fact, the doped CuO2 planes of the $x=1/16$
CNCOC can be divided into \textquotedblleft odd doped CuO
planes\textquotedblright\ and \textquotedblleft even doped CuO
planes\textquotedblright\ and there is a displacement $\mathbf{Q}=4\mathbf{a}
$ (or $4\mathbf{b}$) between them, as indicated in Fig. \ref{fig6}. This
result implies that though a single CuO$_{2}$ exhibits the $4\sqrt{2}a\times
4\sqrt{2}a$ checkerboard pattern at $x=1/16$, the results of STM experiment
on this sample may only show the $4a\times 4a$ pattern as illustrated in
Fig. \ref{fig6}. In other words, the theoretical expectation of $4\sqrt{2}%
a\times 4\sqrt{2}a$ pattern in CNCOC is experimentally unobservable.

In conclusion, the observed checkerboard patterns in hole-doped CNCOC
superconductor have been well explained by the newly developed real-space
spin-parallel pairing and superconducting theory. At $x=1/8$, we show for
the first time the real-space phase competition between the nondispersive $%
4a\times 4a$ checkerboard phase and $8a\times 2c$ superconducting
low-temperature tetragonal vortex phase. At $x=1/16$, we show that the
localized Cooper pairs can organize themselves into the most stable
octahedron phase with a global $4a\times 4a$ checkerboard pattern, while the
theoretical expectation of $4\sqrt{2}a\times 4\sqrt{2}a$ pattern \cite%
{patterson} is formed within each doped CuO$_{2}$ plane.


\begin{thebibliography}{99}
\bibitem{hanaguri} Hanaguri, T. Lupien, C. Kohsaka, Y. Lee, D. -H. Azuma, M.
Takano, M. Takagi, H. \& Davis, J. A `checkerboard' electronic crystal state
in lightly hole-doped Ca$_{2-x}$Na$_{x}$CuO$_{2}$Cl$_{2}$. \textit{Nature}
\textbf{430}, 1001-1005 (2004).

\bibitem{kohsaka} Kohsaka, Y. \textit{et al}. An intrinsic bond-centered
electronic glass with unidirectional domains in underdoped cuprates. \textit{%
Science} \textbf{315}, 1380-1385 (2007).

\bibitem{shen} Shen, K. M. \textit{et al}. Nodal quasiparticles and
antinodal charge ordering in Ca$_{2-x}$Na$_{x}$CuO$_{2}$Cl$_{2}$. \textit{%
Science} \textbf{307}, 901-904 (2005).

\bibitem{vershinin} Vershinin, M. Misra, S. Ono, S. Abe, Y. Ando, Y. \&
Yazdani, A. Local ordering in the pseudogap state of the high-T$_{c}$
superconductor Bi$_{2}$Sr$_{2}$CaCu$_{2}$O$_{8+\delta }$

\textit{Science} \textbf{303}, 1995-1998 (2004).

\bibitem{tranquada} Tranquada, J. M. Sternlieb, B. J. Axe, J. D. Nakamura,
Y. \& Uchida, S. Evidence for stripe correlations of spins and holes in
copper oxide superconductors. \textit{Nature} \textbf{375}, 561-563 (1995).

\bibitem{hdchen1} Chen, H. D. Hu, J. P. Capponi, S. Arrigoni, E. \& Zhang,
S. C. Antiferromagnetism and hole pair checkerboard in the vortex state of
high T$_{c}$ superconductors. \textit{Phys. Rev. Lett.} \textbf{89}, 137004
(2002).

\bibitem{tesanovic} Tesanovic, Z. Charge modulation, spin response, and dual
Hofstadter butterfly in high-T$_{c}$ cuprates. \textit{Phys. Rev. Lett.}
\textbf{93}, 217004 (2004).

\bibitem{kivelson} Kivelson, S. A. Bindloss, I. P. Fradkin, E. Oganesyan, V.
Tranquada, J. M. Kapitulnik, A. \& Howald, C. How to detect fluctuating
stripes in the high-temperature superconductors. \textit{Rev. Mod. Phys}.
\textbf{75}, 1201-1241 (2003).

\bibitem{komiya} Komiya, S. Chen, H. D. Zhang, S. C. \& Ando, Y. Magic
Doping Fractions for High-Temperature Superconductors. \textit{Phys. Rev.
Lett.} \textbf{94}, 207004 (2005).

\bibitem{altman} Altman, E. \& Auerbach, A. Plaquette boson-fermion model of
cuprates. \textit{Phys. Rev. B} \textbf{65}, 104508 (2002).

\bibitem{anderson2} Anderson, P. W. Hall effect in the two-dimensional
Luttinger liquid. \textit{Phys. Rev. Lett.} \textbf{67}, 2092-2094 (1991).

\bibitem{huang1} Huang, X. Q. A Real Space Glue for Cuprate Superconductors.
Preprint at $<$http://arXiv.org/cond-mat/0606177v5$>$ (2006).

\bibitem{huang2} Huang, X. Q. Real Space Coulomb Interaction: A Pairing Glue
for FeAs Superconductors. Preprint at $<$http://arxiv.org/abs/0806.3125$>$
(2008).

\bibitem{huang3} Huang, X. Q. A unified theory of superconductivity.
Preprint at $<$http://arxiv.org/abs/0804.1615$>$ (2008).

\bibitem{kamihara} Kamihara, Y. Watanabe, T. Hirano, M. \& Hosono, H.
Iron-based layered superconductor LaO$_{1-x}$F$_{x}$FeAs ($x$=0.05-0.12)
with $T_{c}=26$ K. \textit{J. Am. Chem. Soc}. \textbf{130}, 3296-3297 (2008).

\bibitem{xfchen} Chen, X. H. Wu, T. Wu, G. Liu, R. H. Chen, H. \& Fang, D.
F. Superconductivity at 43 K in SmFeAsO$_{1-x}$F$_{x}$. \textit{Nature}
\textbf{453}, 761-762 (2008).

\bibitem{seul} Seul, M. Domain shapes and patterns-the phenomenology of
modulated phases. Science \textbf{267}, 476 (1995).

\bibitem{slezak} Slezak, J. A. \textit{et al}. Imaging the impact on cuprate
superconductivity of varying the interatomic distances within individual
crystal unit cells. \textit{PNAS} \textbf{105}, 3203-3208 (2008).

\bibitem{patterson} Patterson, C. H. Small polarons and magnetic antiphase
boundaries in Ca$_{2-x}$Na$_{x}$CuO$_{2}$Cl$_{2}$ ($x$=0.06,0.12): Origin of
striped phases in cuprates. P\textit{hys. Rev. B} \textbf{77}, 094523 (2008).

\bibitem{zhou2} Zhou, F. Hor, P. H. Dong, X. L. Ti, W. X. Xiong, J. W. \&
Zhao, Z. X. Anomalous superconducting properties at magic doping levels in
under-doped La$_{2-x}$Sr$_{x}$CuO$_{4}$ single crystals. \textit{Physica C}
\textbf{408}, 430-433 (2004).
\end{thebibliography}
\end{document}